\documentclass{amsart}
\usepackage{graphicx} 
\usepackage{amssymb, amsmath}
\usepackage{blindtext}
\usepackage{fancyhdr}
\title {P NOT EQUAL TO NP}
\author{\footnotesize Daniel Cardona Delgado\\
   \scriptsize Universidad Nacional de colombia\\
   \MakeLowercase{\footnotesize dcardonade@unal.edu.co}\\
}



\begin{document}

\begin{abstract}
This article finds the answer to the question: for any problem from which a non-deterministic algorithm can be derived which verifies whether an answer is correct or not in polynomial time (complexity class NP), is it possible to create an algorithm that finds the right answer to the problem in polynomial time (complexity class P)? For this purpose, this article shows a decision problem and analyzes it to demonstrate that this problem does not belong to the complexity class P, but it belongs to the class NP; doing so it will be proved that it exists at least one problem that belongs to class NP but not to class P, which means that this article will prove that not all NP problems are P.

\end{abstract}

\maketitle

\section{Introduction}\label{section}

The "P versus NP" problem proposed by Stephen Cook and Leonid Levin is a question that has persisted in the computing ambit since 1971. It is formulated as follows: Is every language accepted by some non-deterministic algorithm in polynomial time also accepted by some deterministic algorithm in polynomial time? \cite{cook2003importance}, or equivalently: For all decision problems from which an algorithm can be derived that checks whether an answer is correct or incorrect in a polynomial number of steps (NP complexity problems), it is also possible to find an algorithm that generates a correct answer in a polynomial number of steps (complexity problems P)?, with the algorithm of P being executed by a Turing machine and that of NP by a non-deterministic Turing machine \cite{martin2004introduction}. It should be noted that NP can also be defined equivalently as the set of problems in which the result can be verified in polynomial time \cite{sipser2013introduction}. This problem is of great importance to the world of computing, mainly because of the repercussion that it would have in cryptographic security (if P=NP would have a quick solution, the security of the protected information will be affected because it is based mainly on NP problems). This problem is called one of the 7 problems of the millennium. Many mathematicians believe that $P \neq NP$, although no proof of this has been found. The idea of this article is focused on creating a problem that is NP, but not P; to do this, the proposed problem will be analyzed through logic proving that it is not P. Later, it will be proved that it is NP by creating an algorithm that checks if an answer is correct or incorrect in polynomial time for the problem created.

\section{Methods}

This section shows the definition of a decision problem that is NP. For this porpuse a game will be created which is inspired by Conway's game of life and its turing universality \cite{rendell2002turing}. The game here created, which was called "Debilandia", is a generations based game (the game state changes based on generation) and is structured on a grid composed of 4x4 sub-grids which are called Tiles.\\ 

The tiles in the game cannot overlap and are composed of localized points. In each generation the content of the grid changes due to different effects that depend on the tiles already located on the grid. A tile also belongs to a type: tip, rule or tape, and can represent a constant value. Furthermore, a group of tiles of type "rule" can be assigned to a "Rules set".\\
  
Each tile used in the game is explained below, where each "X" in the figure represents the location of a point in the sub-grid that conforms the tile:\\ \\

\begin{itemize}
    \item[1] "Tip": This tile is of type "tip" and should have a tile with a "reading value" directly above it, which in turn has a "status tile" directly above it, which will be considered the status of the tip $S$  (figure 1)
    \item[2] "Tape value 1": It is considered a constant of value 1 and it is of type "tape" (figure 2(a))
    \item[3] "Tape value 0": It is considered a constant of value 0 and it is of type "tape". (figure 2(b))
    \item[4] "Read value 1": It is a "rule" type tile. It is identified as $R_1$ when it belongs to a set of rules and represents a value of 1. (figure 3(a))
    \item[5] "Read value 0": It is a "rule" type tile. It is identified as $R_1$ when it belongs to a set of rules and represents a value of 0. (figure 3(b))
    \item[6] "Status value 1": It is a "rule" type tile. It is identified as $R_2$ when it belongs to a set of rules and represents a value of 1. (figure 4(a))
    \item[7] "Status value 0": It is a "rule" type tile. It is identified as $R_2$ when it belongs to a set of rules and represents a value of 0. (figure 4(b))
    \item[8] "Write 1": It is a "rule" type tile. It is identified as $R_3$ when it belongs to a set of rules and represents a value of 1.(figure 5(a))
    \item[9] "Write 0": It is a "rule" type tile. It is identified as $R_3$ when it belongs to a set of rules and represents a value of 0. (figure 5(b))
    \item[10] "Change status to 1": It is a "rule" type tile. It is identified as $R_4$ when it belongs to a set of rules and represents a value of 1. (figure 6(a))
    \item[11] "Change status to 0": It is a "rule" type tile. It is identified as $R_4$ when it belongs to a set of rules and represents a value of 0. (figure 6(b))
    \item[12] "Movement 1": It is a "rule" type tile. It is identified as $R_5$ when it belongs to a set of rules and represents moving the tape to the left. (figure 7(a))
    \item[13] "Movement 0": It is a "rule" type tile. It is identified as $R_5$ when it belongs to a set of rules and represents moving the tape to the right. (figure 7(b))
\end{itemize}

\begin{figure}[htb!]
\centering
\includegraphics[scale=0.7]{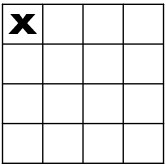}
\caption{}
\end{figure} 

\begin{figure}[htb!]
\centering
\includegraphics[scale=0.75]{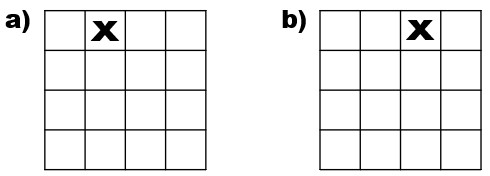}
\caption{}
\end{figure} 

\begin{figure}[htb!]
\centering
\includegraphics[scale=0.75]{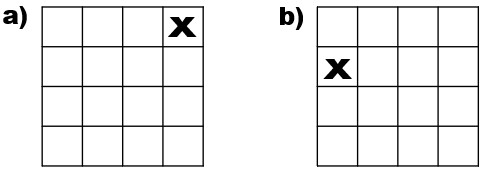}
\caption{}
\end{figure} 

\begin{figure}[htb!]
\centering
\includegraphics[scale=0.75]{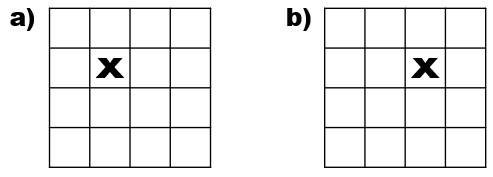}
\caption{}
\end{figure} 

\begin{figure}[htb!]
\centering
\includegraphics[scale=0.75]{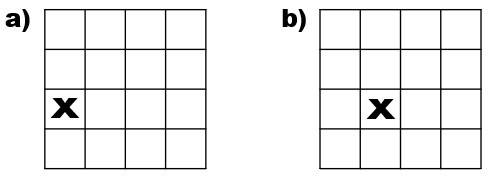}
\caption{}
\end{figure} 

\begin{figure}[htb!]
\centering
\includegraphics[scale=0.75]{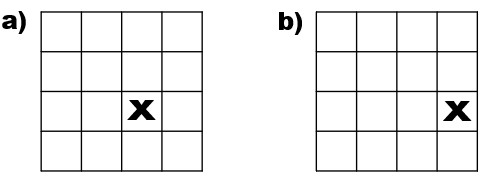}
\caption{}
\end{figure} 

\begin{figure}[htb!]
\centering
\includegraphics[scale=0.75]{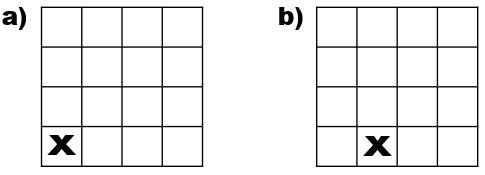}
\caption{}
\end{figure} 

\newpage

The tip, at the beginning of each generation reads the tile directly below it, which should be a tape-type or a rule-type tile:\\

	If the tile it reads is a Tape type tile, it will create a new read tile just above the tip (overwriting any existing tile) of Q value, where Q is the value of the readed tape-type tile; afterwards it checks the tile that is on the top right of the tip, which we will call $R_1$, and the 4 tiles to the right of $R_1$, one by one from left to right ($R_2, R_3, R_4, R_5$). These set of values will represent a packet of rules (or instructions), being $R_1$ the read value, $R_2$ the state, $R_3$ the value to be written, $R_4$ the change of state and $R_5$ the movement. If the read value Q is equal to the value of $R_1$ and the value of the tip state $S$ is equal to $R_2$, then the value of $R_3$ is written on the tile directly below the tip, as a tape tile of the same value as $R_3$. Then it places a status token of the value of $R_4$ on the tip state tile and moves all tape tiles in the direction indicated by the value of $R_5$. If the conditions (the $R_1$ and $R_2$ values) of that rule packet are not met, it searches directly above the current rule packet looking for another packet that met the conditions, and repeats the process until the conditions are met. \\   

    The tape is moved based on the value of $R_5$ as follows: all the tiles that are in the row below the tip are pushed towards the direction indicated by $R_5$ (each tile moves only one sub-grid space at a time). After having carried out the movement indicated by $R_5$, the generation is considered to be over.\\
    
    If the tip reads a rule-type tile, it creates a copy of the tile (of the same type and value) in the highest incomplete rule packet that appears on top of the tip. Being an incomplete packet, a packet that still doesn't have the tiles all of the tiles that conforms the packet: $R_1, R_2, R_3, R_4$ or $R_5$. The new tile will be created in order, that is, the tile $R_1$ must always be in the right column of the tip, followed in the same row by the tiles $R_2$, $R_3$, $R_4$ and $R_5$, in that order, without repetition. After the new tile is created, all the tiles that were to the left of the just read tile are moved one sub-grid to the right (replacing the value just read). After carrying out the movement just explained, the generation is considered to be over.\\
    
    If the tip cannot execute any of the actions mentioned of if there is more than one tip in the initial configuration of the grid, the game is considered invalid and over.

\subsection{Problem}

The problem statement created is the following: given a set A of k positive integers different from 4, 25, 43, 7, 5, 2, with k being a random value, and given a set B that is the union of the set A with the values 4, 25, 43, 7, 5, 2. Determine if there exists a list L, which is a variation with repetition of the set B and represents the points of an initial state of the game created "Debilandia" and that generates a Turing machine with all its rules $R_1,R_2,R_3,R_4,R_5$ already established,  under the specific conditions listed below:   

    \begin{enumerate}
        \item The first number of L must be a 2 value.
        \item After the value 2 there must be a pair of numbers $P_0$ that must belong to set A (since the intersection between set A and set B is the same set A) then there must be a 7 and another pair of numbers $P_1$ that must belong to the intersection between set A and set B, and again a 7 and so on. 
        \item The tuples $P_0$, $P_1$, $P_2$ and so on, must not have repeated values and must correspond to all the possible tuples generated from a variation with repetition of two elements from the set A.
        \item After the last tuple there must be a 5.
        \item After that, a 4 value must appear for every generation the game must make before stopping.
        \item The initial state of the game is generated assuming that each of the tuples $P_i$ is the coordinate of a point in the general grid of the game.
        \item Finally, after the sequence of all 4s, a 25 value must appear if the game with that configuration spawns a Turing machine that stalls or a 43 if it doesn't.
    \end{enumerate}

\subsection{Demonstration that the game created "Debilandia" is Turing complete }

A Turing machine can be built inside the game because "Debilandia" is composed of:

    \begin{enumerate}
        \item A straight line of tape-like tiles that would represent the memory (with values of 1 and 0); with a type alphabet $\sum = {0,1}$
        \item One or more packets of rule-type tiles (to the right above the tip), which will be the rules that will tell the tip what to do. It makes use of the same alphabet $\sum$ as the tape-type tiles.
        \item An initial state.
        \item A reading tip ("The Tip") above the tape. The tip can move the tape left and right, read values on the tape, write values to the tape or change the state of the machine according a set of rules. 
    \end{enumerate}
    
It can be said that it is Turing complete, since it can run a universal Turing machine. In this case, it would be the same machine mentioned above, only now, the rules are included at the beginning of the tape using rules tokens, which, when read by the tip, are added to the rules in the grid; Therefore, this Turing machine can simulate any other Turing machine.

\subsection{The problem is not P}

This problem would not be P, because when using the game "Debilanda", in order to get a right answer, the problem puts as a condition the generation of a Turing Machine inside the game "Debilandia", for this reason, this problem is reducible to the "halting problem", proposed by Turing, who proved that Turing machines and precisely that the halting problem was undecidable \cite{turing1936computable}. So, since the problem proposed in this article can be reduced to the "halting problem", it is also undecidable.\\

Based on the fact that the problem presented is undecidable, it is known that you must go through the entire game process to know what the final state will be. Besides, because the last number of the list L forces the final state of the game to be known, it is necessary for any algorithm to execute all the generations of the game to find that number. In order to do this, all the cells required by L must be placed, that is, all the Pi tuples that are part of the same L.\\

In consequence, an algorithm that finds a right answer to the prbolem must put $(M!)^2$ cells in the game grid, where M is the number of numbers in the set A, which is an exponential number based on the size of the set A. Having to apply the rules of the game to all cells (which will be part of the tiles) to discover the last number of L, an algorithm that finds a right  answer to the problem must perform at least that ammount of operations ($(M!)^2$).\\

Because set A is contained in set B, an algorithm that finds an answer only requires the values of set B to generate an answer. Given that the cardianlity of B is M + 6, the size of the input for an algorithm N that finds a correct answer to this problem is M + 6.\\

It is worth mentioning that M! is an exponential operation with respect to M, which is clearly seen by looking at the way $M!$ grows as M grows. Stirling's formula \cite{bagui2013nonrigorous} also shows the growth of the factorial using limits:\\

\begin{center}
\begin{equation}
\lim_{n \to \infty} \frac{(n^n)(e^{-n})(\sqrt[2]{2 \Pi n})}{n!} = 1 \qquad\qquad
\end{equation}
\end{center}

From this equation we can define the value to which n! tends by means of limits, like this:\\

\begin{center}
\begin{equation}
n! \thickapprox (n^n)(e^{-n})(\sqrt[2]{2 \Pi n}) \qquad\qquad
\end{equation}
\end{center}

Operation of which it can be affirmed that n! is exponential in its limit (it is worth mentioning that the fact that $e^{-n}$ appears does not take away from the fact that it is exponential , since e is a constant, which means that the $n^n$ in the long term influences the growth form of the operation much more than the term $e^{-n}$); with all this, it makes it clear that n! is an exponential operation with respect to n, at a practical and theoretical level. \\

For all this it is correct to affirm that an algorithm that finds a correct answer to the proposed problem must be at least exponential with respect to the size of the input N, since $N!-6$ is exponential with respect to N. Therefore, the problem created cannot be solved in a polynomial number of steps with respect to the input N and therefore the problem created does not belong to the complexity class P.

\subsection{the problem is NP}

Now it will be demostrated that the problem treated belongs to the complexity class NP. For this, it is taken into account that NP is the set of problems of which, there is an algorithm that verifies if the answer is correct or incorrect in polynomial time. 

Faced with this situation, an algorithm f(n) was created to verify the answer for the problem of this article. It is worth mentioning that when calculating how many steps an algorithm performs, the results may vary depending on the language used; variations that will be ignored because they are too small to represent a change at a practical level. The number of steps can also vary by the initial organization of points since there may be many points left that do not generate valid tiles and many tiles that are not part of the specification of the Turing machine (in a correct answer).  \\

For the verificaion algorithm you only must consider:
\begin{enumerate}
        \item T= number of pairs with repetition that the answer has. 
        \item P = 2T 
        \item E= generations that the game must complete until it stabilizes with the given configuration (considering stabilizing to enter a state without changes or enter a cycle that repeats itself in exactly the same way every time); that is, the number of numbers 4 in a row after 5 and before a 25 or a 43.
        \item N = P + E + T + 4 = number of input numbers to the algorithm.
\end{enumerate}

The process that the algorithm will follow is showed bellow. In brackets appears the number of steps that the algorithm needs to make each step, taking into account that when there are different options, the number of steps will be taken to the longest branch; that is, the maximum number of steps that could be done:

    \begin{enumerate}
        \item Choose the first position in the input list L, which must be a "2", if it is not, it shows that the answer is wrong. (1)
        \item Then it starts traversing the list storing the following numbers in a list $Q_i$ until it finds a 7, at that point it creates another list $Q_{i+1}$; the $Q_i$ list must have exactly 2 numbers, if it doesn't have two elements, it returns that the answer is not correct. 
        \item Step two is repeated creating lists $Q_i$ until a 5 is found. (2P+T+1).
        \item Then it locate points int the grid using the $Q_i$  lists as coordinates and divide the grid into 4x4 size squares (from the bottom left point) and detect the tiles that have been created. (P+ 17T).
     	\item Subsequently, it will continue advancing in the list, after the 5 value, counting how many 4s there are before it meets a 25 or a 43. If a number that is not a 4 appears before a 25 or a 43, it will return that the answer is wrong. Then it will run on the game grid, previously created, the same number of generations as numbers 4 has been read and it will check if the last generation of the game stops or not. (2E+ET+1+2) 
        \item At the same time that it performs the generations, it checks if it is true that the game has generated a Turing machine, detecting the 3 basic parts of a Turing machine, which are the tape, the tip and the rules (with the specifications of section 2.1) if it doesn't find them, it shows that the answer is wrong. In order to do this, at the first generation of the game, it looks for a straight line of tape-like tiles with a tip tile at the top row. (ET) 
        \item Finally, it reads one more number further down the list, detects if the number is a 25 or a 43, and checks to see if the last build you ran the game stopped or not. (2)
        \item If the last number was 25 and the Turing machine stopped in the last generation executed, it considers the problem has a right answer, or if the number was 43 and the Turing machine does not stop it consider it also a right answer, but if this is not fulfilled or there is a number after the 25 or 43, it shows that the answer is wrong. (2) 
    \end{enumerate}

\subsection{Number of steps}

Subsequently, each number of steps is added to conclude that the algorithm needs the following number of steps to demonstrate whether an answer is correct or not:\\

\begin{center}
\begin{equation}
f(N) = 2ET + 2E + 3P + 18T + 10 \qquad\qquad
\end{equation}
\end{center}

\subsection{Proof that the algorithm is polynomial}
To demonstrate that the algorithm takes a polynomial time with respect to the input N, it is shown that equation (3) is less than or equal to a polynomial function based on N, taking advantage of the fact that for practical purposes N, P, E and T are all positive integers (your input will never be a negative quantity), then is possible to find a formula mathematically greater than or equal to the original function, repeating the process with a new equation until arriving at a formula expressed only in N: 

\begin{center}
\begin{equation}
 f(N)\leq 2ET + 33N \qquad\qquad
\end{equation} 
\begin{equation}
 f(N)\leq 2N^2+ 33N \qquad\qquad
\end{equation}
\end{center}

Considering the ecuation (5), it is possible to say that the algorithm decides if the answer is correct or incorrect in equal or less than $2N^2 + 33N$ steps, which is a polynomial function with respect to the original N. Knowing this, it is correct to affirm that the presented algorithm belongs to NP, since it can be demonstrated whether an answer is correct or incorrect in, at most, a polynomial number of steps with respect to the input N.

\section{Results}

The union of all the information collected throughout this document led us to the main result and objective of this article, which was to discover if P is equal to NP. The proposed problem belongs to the complexity class P, since it is impossible to create an algorithm that finds an answer in polynomial time, and there is also an algorithm that can tell if an answer to this problem is correct or not in polynomial time (so the problem would belong to the complexity class NP). By determining that the problem belongs to the complexity class NP, but not to the complexity class P, it is shown that there is at least one problem that is NP, but is not P, from which it is concluded that not all NP problems are P, hence $P \neq NP$.

\section{Discussion and Conclusions}

The conclusion of this article is the one previously exposed: that $P \neq NP$. It means that for NP-Complete problems (which are a set of problems that can be reduced in polynomial time to any other NP problem) there is no algorithm that can find an answer in polynomial time, because if there existed at least one NP-Complete problem that was also P then all NP problems would be P, which is not the case. Other than this, there is no clear path or future to which the proof that $P \neq NP$ can lead.

\bibliographystyle{amsplain}
\bibliography{Article_P_NP}

\providecommand{\bysame}{\leavevmode\hbox to3em{\hrulefill}\thinspace}
\providecommand{\MR}{\relax\ifhmode\unskip\space\fi MR }
\providecommand{\MRhref}[2]{%
  \href{http://www.ams.org/mathscinet-getitem?mr=#1}{#2}
}
\providecommand{\href}[2]{#2}
\begin{thebibliography}{1}

\bibitem{bagui2013nonrigorous}
Subhash~C Bagui, Sikha Bagui, and Rohan Hemasinha, \emph{Nonrigorous proofs of
  stirling's formula}, Mathematics and computer education \textbf{47} (2013),
  no.~2, 115.

\bibitem{cook2003importance}
Stephen Cook, \emph{The importance of the p versus np question}, Journal of the
  ACM (JACM) \textbf{50} (2003), no.~1, 27--29.

\bibitem{martin2004introduction}
John~C Martin, Jorge~Luis Blanco~y Correa~Magallanes, et~al.,
  \emph{Introduction to languages and the theory of computation. lenguajes
  formales y teor{\'\i}a de la computaci{\'o}n}, 2004.

\bibitem{rendell2002turing}
Paul Rendell, \emph{Turing universality of the game of life}, Collision-based
  computing (2002), 513--539.

\bibitem{sipser2013introduction}
M~Sipser, \emph{Introduction to the theory of computation. 3th}, Cengage
  Learning (2013).

\bibitem{turing1936computable}
Alan~Mathison Turing et~al., \emph{On computable numbers, with an application
  to the entscheidungsproblem}, J. of Math \textbf{58} (1936), no.~345-363, 5.

\end{thebibliography}
\end{document}